\documentclass[a4paper,11pt]{article}

\pdfoutput=1

\usepackage{jcappub}
\usepackage{amsmath}
\usepackage{amsfonts}
\usepackage{amssymb}
\usepackage{mathtools}
\usepackage{graphics}
\usepackage{citesort}
\usepackage{graphicx}
 \usepackage{url}
 \usepackage{xcolor}
\usepackage{soul}
\usepackage[utf8]{inputenc}

\usepackage{hyperref}
\usepackage{color}

\title{A closer look at the $pp$-chain reaction in the Sun: Constraining the coupling of light mediators to protons}

\author[a]{Anna M. Suliga,}
\author[a]{Shashank Shalgar,}
\author[b]{and George M. Fuller}
\affiliation[a]{Niels Bohr International Academy and DARK, Niels Bohr Institute,
University of Copenhagen, Blegdamsvej 17, 2100, Copenhagen, Denmark}
\affiliation[b]{Department of Physics, University of California, San Diego, La Jolla, CA 92093-0319, USA}

\emailAdd{anna.suliga@nbi.ku.dk}
\emailAdd{shashank.shalgar@nbi.ku.dk}
\emailAdd{gfuller@physics.ucsd.edu}

\abstract{
The $pp$-chain of nuclear reactions is the primary route for energy production in the Sun. The first step in that reaction sequence converts two protons to a deuterium nucleus with the emission of a positron and electron neutrino. This reaction is extremely slow because it is a weak interaction, and significantly, it involves quantum tunneling through the Coulomb barrier. Though the reaction rate can be calculated with high confidence in the Standard Model, it has not been measured at solar energies. If there exist interactions that are engendered by non-standard mediators then the rate of this reaction in the Sun could be altered.
We probe such non-standard interactions by comparing calculations of solar evolution to the current solar system age in the presence and absence of the non-standard mediators. These reveal ranges of non-standard mediator mass and couplings that are inconsistent with measured properties of the Sun, including solar neutrino results. Our constraints on these non-standard parameters, in many cases overlapping those derived via other considerations, could be extended further with better confidence in the value of the metalicity of the Sun and the solar neutrino CNO flux.
Intriguingly, our work reveals a degeneracy between the solar metalicity and the presence of the invoked non-standard mediators.
}

\begin{document}

\maketitle
\flushbottom

\section{Introduction}
\label{sec:intro}
In this paper we explore a surprising connection between speculative issues in elementary particle physics and the Sun. In particular, we show how the extreme sensitivity of nuclear reaction Coulomb barrier penetration at the low energies of the solar core could be leveraged to probe aspects of the non-standard interactions involving light mediators recently invoked to explain anomalies in short baseline neutrino experiments \cite{Liao:2018mbg, Bertuzzo:2018itn, Ballett:2018ynz, Arguelles:2018mtc,Denton:2018dqq,Datta:2020auq}.

The Standard Solar Model (SSM) has been extremely successful in revealing the inner workings of the Sun. It has enabled using solar neutrino measurements to probe fundamental neutrino physics. In fact, that effort played a key role in the discovery of neutrino flavor oscillations and establishing non-vanishing neutrinos masses. This is the only concrete discovery to date of Beyond Standard Model (BSM) physics. The SSM has also played a vital part in constraining BSM scenarios \cite{Schlattl:1998fz,Beacom:1999wx,Beacom:2002cb,Raffelt2008,Gondolo:2008dd,Redondo:2013lna,Vinyoles_2015,Vinyoles:2015khy,Hardy:2016kme,Giannotti:2017hny,Aguilar-Arevalo:2016zop,Huang:2018nxj,AristizabalSierra:2020zod,Fabbrichesi:2020wbt,Alonso-Alvarez:2020cdv,Caputo:2020quz,OHare:2020wum,Guarini:2020hps}. Most of these studies rely on energy loss arguments \cite{Raffelt:1996wa}. Such arguments revolve around the creation of light particles escaping the core of the Sun, thereby altering energy transport and so modifying the solar temperature and other solar structure parameters in ways amenable to constraint. 
In this paper, we focus instead on a different BSM physics possibility: modified interactions between quarks/nucleons. 

The conversion of four protons to an alpha particle is the main route for energy generation in main sequence stars like the Sun. Obviously, this process requires the weak interaction to convert two of the protons into neutrons. In the Sun and other lower main sequence stars, this is accomplished primarily by the first reaction step in the $pp$-chain, the fusion of two protons to produce a deuteron (D), positron, and electron flavor neutrino. This first step, $p+p\rightarrow \mathrm{D} + e^{+} + \nu_{e}$, is a weak nuclear reaction. It is extremely slow in the Sun for two reasons. 
First, the process is a weak interaction, and second, the temperature at the center of the Sun is not high enough to allow for the classical penetration of the Coulomb barrier and the reaction must proceed by quantum tunneling through the Coulomb barrier. The probability that a proton will undergo the first step of the $pp$-chain reaction is one in 10 billion years, which gives the approximate time-scale of the life-span of the Sun. { The subsequent reactions in the $pp$-chain, which involve the Coulomb barrier penetration, are mediated by the strong or electromagnetic forces. They occur on much faster timescales. This fact highlights the importance of the key first step.}
This key first step has {also} not been measured in any laboratory. On the other hand, all other $pp$-chain and many other nuclear reactions of consequence in astrophysics have been measured in the laboratory. { 

Reaction cross section data usually is obtained in energy ranges well above the range of most effective stellar energies, necessitating extrapolation of the cross section to the relevant low energy regime. This extrapolation is effected via an S-factor, factoring out a generic barrier penetration and geometric factor. Sometimes this procedure is augmented with a cross section-energy dependence taken from a nuclear model. For example, this latter procedure is followed in cases for light nuclear reactions that proceed through the wings of a higher energy resonance or resonances. Introducing BSM modifications to the effective Coulomb barrier is, as we will show below, a tricky complication. Though in principle cross sections extrapolated from laboratory measurements would already include the effects of this BSM physics, in practice this new physics may only manifest at very short range, calling into question the extrapolation procedure. In fact, as we will see, the current experimental precision on the relevant cross sections is not sufficient to rule out the very small BSM barrier modifications that we consider here. However, this level of BSM physics modification of the barrier {\it would} have a significant effect on the weak interaction first step in the $pp$-chain.}

Absent experimental measurements, the SSM must rely on a theoretical calculation of the rate for the first step of the $pp$-chain reaction. With Standard Model physics this can be done with high confidence \cite{1994ApJ...420..884K,1968ApJ...152L..17B,2003PhRvC..68e5801B,1990ApJ...359L..67G}. In fact, effective field theory and {\it ab initio} techniques (see {Refs.} \cite{Savage:2000hs} and \cite{2002CzJPh..52B..49P}) can be brought to bear on calculations of the nuclear matrix element for this and other light nuclei 
\cite{2001PhLB..520...87B,2002nucl.th...7008B,2002PhLB..549...26B,2001PhRvC..64d4002K,2011RvMP...83..195A,1994ApJ...420..884K,1968ApJ...152L..17B,2003PhRvC..67e5206P}, further increasing confidence in the veracity of the Standard Model result. Moreover, the reverse process can be measured \cite{Pizzone:2014xrw,Tumino_2014}, albeit at energies considerably higher than those relevant for solar fusion.

However, the presence of interactions beyond the Standard Model could modify the matrix element and the overall reaction rate at solar core energies. 
In this work, we calculate the evolution of the Sun with such BSM modifications to the rate of $p+p\rightarrow \mathrm{D} + e^{+} + \nu_{e}$. {  We also discuss BSM modifications to other reactions relevant to the $pp$-chain and CNO cycle.} In this way, we can probe a range of parameters in a general framework for non-standard interactions in the $p+p$ channel.

We consider massive non-standard vector boson and scalar mediators that can couple to nucleons. 
{ There would be an additional repulsive force between the protons in the case of a vector boson mediator, and an attractive force for the scalar, respectively decreasing and enhancing the quantum tunneling probability at a given temperature.} This would alter the evolution and current structure of the Sun.
For example, attaining hydrodynamic equilibrium in the presence of BSM barrier penetration modification would require the center of the Sun to have a different temperature, with a consequent change in the rate of the subdominant Carbon-Nitrogen-Oxygen (CNO) cycle contribution to energy generation and neutrino emission. We use a 1D stellar evolution/hydrodynamics code to calculate the effect of a massive non-standard vector or scalar boson on the evolution and structure of the Sun, and use the results to { probe or constrain} the couplings and the masses of these BSM particles. We want to stress that the main goal of this work is to{ explore the effects of the non-standard mediators coupling to protons on the solar evolution and to derive limits on the mass and coupling of non-standard mediators where possible. Our intent in exploring this hypothetical BSM physics is not to use it to solve the solar metallicity problem~\cite{Antia:2005mg} or any other potential problem with the current solar model.}

This paper is organized as follows. In Sec.~\ref{sec:Non-Standard Interactions}, we discuss the motivation for considering models of non-standard interactions involving a massive vector boson. In Sec.~\ref{sec:Standard Solar Model and beyond} we briefly review the SSM with a focus on issues pertinent to this paper. In Secs.~\ref{sec:Effective dependence on the new coupling} and \ref{sec:Results} we describe the numerical techniques used in this paper and present the results. We discuss the implications of these results and put them in a broader context in Sec.~\ref{sec:Discussion}. Finally, we conclude in Sec.~\ref{sec:Conclusions}. 

\section{Non-Standard Interactions}
\label{sec:Non-Standard Interactions}

The existence of Dark matter and non-vanishing neutrino masses indicate physics beyond the Standard Model. Furthermore, the smallness of the neutrino masses suggests that these are suppressed by a high energy scale, which is non-standard in origin. The existence of non-standard interactions could potentially have implications for low energy interactions, although none have been found. The most probable scenario for non-standard physics involves additional gauge interactions, which would imply the existence of gauge bosons beyond the ones currently known. Interactions of these non-standard gauge bosons with Standard Model particles cannot be ruled out. 

Physics beyond the Standard Model has been invoked to explain various anomalies~\cite{Aguilar:2001ty, Aguilar-Arevalo:2018gpe, Serebrov:2017bzo, Pattie:2017vsj, Krasznahorkay:2015iga}. Although the origin of these anomalies is not known, and they could very well be due to experimental uncertainties, efforts have been made to address these anomalies by invoking interactions that are beyond the Standard Model. This includes the LSND/MiniBoone anomaly~\cite{Liao:2018mbg, Bertuzzo:2018itn, Ballett:2018ynz, Arguelles:2018mtc,Denton:2018dqq,Datta:2020auq}, the neutron decay anomaly~\cite{Fornal:2018eol, Barducci:2018rlx, Tang:2018eln}, the Beryllium-8 decay anomaly~\cite{Feng:2016ysn, Banerjee:2018vgk, Dror:2017ehi, Kahn:2016vjr}, and the KOTO anomaly \cite{Egana-Ugrinovic:2019wzj}. There are several studies in the literature dedicated to constraining the interaction of Standard Model particles with the non-standard sector. Because neutrinos interact with other particles and with themselves via the weak interaction, the non-standard interactions of neutrinos is the least constrained and, consequently, the most interesting. 

The interactions of charged leptons are measured with much better accuracy than are the interactions of nucleons \cite{Tanabashi:2018oca}. 
Specifically, electroweak precision measurements constrain the interactions of the charged leptons to a greater level of confidence than do experimental probes of the hadronic sector. For example, the presence of any non-standard mediator that interacts with charged leptons would also inevitably modify the magnetic moment of the muon, vitiating the ultra-precise agreement between standard model theory and experiment~\cite{Czarnecki:2001pv}. 

In this paper, we focus on the non-standard interactions that could plausibly affect the nucleons. { The participation of nucleons in non-standard interactions is of particular importance. For example, such interactions could encompass} modification of the neutrino-nucleon interactions by 6-dimensional effective operators of the form, 
\begin{eqnarray}
\mathcal{O}^{p\nu}_{6} \sim \langle \bar{\nu} | \Gamma^{\mu} | \nu \rangle \langle \bar{p} | \Gamma_{\mu} | p \rangle,\\
\mathcal{O}^{n\nu}_{6} \sim \langle \bar{\nu} | \Gamma^{\mu} | \nu \rangle \langle \bar{n} | \Gamma_{\mu} | n \rangle,
\label{6dim}
\end{eqnarray}
where, $p$ and $n$ denote the proton and neutron, respectively, while $\Gamma^{\mu}$ is a linear combination of $1, \gamma^{5}, \gamma^{\mu}, \gamma^{\mu}\gamma^{5}, \frac{i}{2}[\gamma^{\mu},\gamma^{\nu}]$, with $\gamma^{\mu}, \gamma^{\nu}, \gamma^{5}$ denoting the Dirac-gamma matrices. 
Those operators can arise in several models. The simplest are theories with an additional U(1) gauge symmetry. These are invoked for the neutrino non-standard interactions in, for example,~Refs.~{\cite{Harnik:2012ni,Farzan:2016wym,Cerdeno:2016sfi,Denton:2018xmq,Croon:2020lrf,Coloma:2020gfv,Suliga:2020jfa,Flores:2020lji}}.

The most natural way to incorporate the 6-dimensional operators described in eq.~\ref{6dim} is to add a new particle that mediates an interaction between neutrinos and nucleons. The new interaction leads to resonant enhancement of cross-section near the center-of-mass energies that are close to the mass of the new mediator. The new mediator can also lead to new channels of decay for particles in the Standard Model. 

However, if neutrinos interact with nucleons by exchanging a mediator, it is plausible, likely inevitable, that similarly neutrinos interact with themselves, and likewise for nucleons. We likely cannot avoid having 6-dimensional operators of the form,
\begin{eqnarray}
\mathcal{O}^{pp}_{6} \sim \langle \bar{p} | \Gamma^{\mu} | p \rangle \langle \bar{p} | \Gamma_{\mu} | p \rangle.
\label{pp6dim}
\end{eqnarray}
We focus on the protons instead of neutrons in this paper, though we will assume that whatever interactions we posit are isospin invariant and parity preserving. 

 It should be noted that the operator in eq.~\ref{pp6dim} is written as an interaction between protons and not quarks; this is on purpose, and for two reasons. The operator in eq.~\ref{pp6dim} can arise due to the interaction of a non-standard mediator with the constituents of the proton, but since it is an operator in an effective field theory, the constraints on the coupling strength of the operator cannot be extrapolated from the interactions with other composite particles. The process of extrapolation inherently assumes that we can ascertain the interaction strength of the new mediator with each constituent of every composite particles in question.

We should also be careful when comparing constraints from two different experiments or observations when dealing with low energy effective operators. The low energy effective operator is presumably a result of an UV-complete theory and hence has a well-defined renormalization group evolution of the coupling strength. When dealing with specific operators of the form eq.~\ref{pp6dim} in isolation, the constraints at one energy scale may not be valid at another energy scale. It is therefore prudent to ensure that constraints on low energy effective operators are not unreasonably extrapolated from one scenario to another and a multiplicity of constraints from several experiments and observations is vital.

Extrapolation of constraints from one scenario to another is, of course, more reasonable in certain scenarios than in others. In the case of invoking a new mediator to explain the low energy excess in LSND/MiniBoone~\cite{Liao:2018mbg, Bertuzzo:2018itn, Ballett:2018ynz, Arguelles:2018mtc,Denton:2018dqq,Datta:2020auq}, an additional interaction between nucleons would be virtually inevitable. A non-standard interaction that can explain the LSND/MiniBoone excess involves two couplings: the coupling constant of the non-standard mediator to neutrinos; and the coupling constant of the non-standard mediator to quarks. A sufficiently large coupling of the non-standard interaction to quarks would be in conflict with the SSM. Additionally, the recoil experienced by nucleons inside the carbon nuclei involves relatively low energies. In this paper, we focus on the constraints on additional interaction between protons at low energy in the Sun. The constrains we obtain are therefore of potential importance to the models that invoke an additional mediator to explain the excess events in LSND/MiniBoone.

Operators of the form eq.~\ref{pp6dim}, cannot lead to creation or annihilation of particles; they cannot contribute to the decay of particles with low masses. However, these operators would result in an attractive or repulsive force between protons depending on the form of the matrix $\Gamma^{\mu}$. We consider two representative cases in this paper: $\Gamma^{\mu}=\gamma^{\mu}$ and $\Gamma^{\mu}=1$, corresponding to a vector and scalar mediator, respectively. 
The vector mediator, denoted by $Z^{\prime}$, results in a repulsive force between protons, while the scalar mediator, denoted by $\phi$, results in an attractive force. Of course, these interactions would be in addition to the usual nucleon-nucleon standard model strong interaction and to the Coulomb repulsion between the protons~\cite{Zee:706825}. The interaction between the new mediators and the protons is fully described by the corresponding interaction Lagrangians,
\begin{eqnarray}
\mathcal{L}^{Z^\prime} = g Z^\prime_\mu \bar{\mathrm{p}} \gamma^{\mu} \mathrm{p} \ ,
\mathcal{L}^{\phi} = g \phi \bar{\mathrm{p}}  \mathrm{p} \ .
\end{eqnarray}
where $g$ is the coupling to the new mediator. 

The new non-standard interaction (NSI) leads to an additional repulsive or attractive potential between the protons of the form,
\begin{equation}
V(r) = \frac{e^{2}}{r} \pm \frac{g^{2}}{r} e^{-m_{\{Z^\prime, \phi\}}r},
\label{nsipot}
\end{equation} 
where, $e$ is the Standard Model coupling, $e^2\approx 1/137.036$, $m_{(Z^\prime,\phi)}$ is the mass of the appropriate mediator, and $g$ is the coupling of the mediator to the proton. 
The range of the NSI between the protons may be larger than that of the nucleon-nucleon strong potential, and so may modify Coulomb barrier penetration, from the classical turning point to the proton-proton contact point.

\begin{figure}[t]
 \centering
 \includegraphics[scale=0.41]{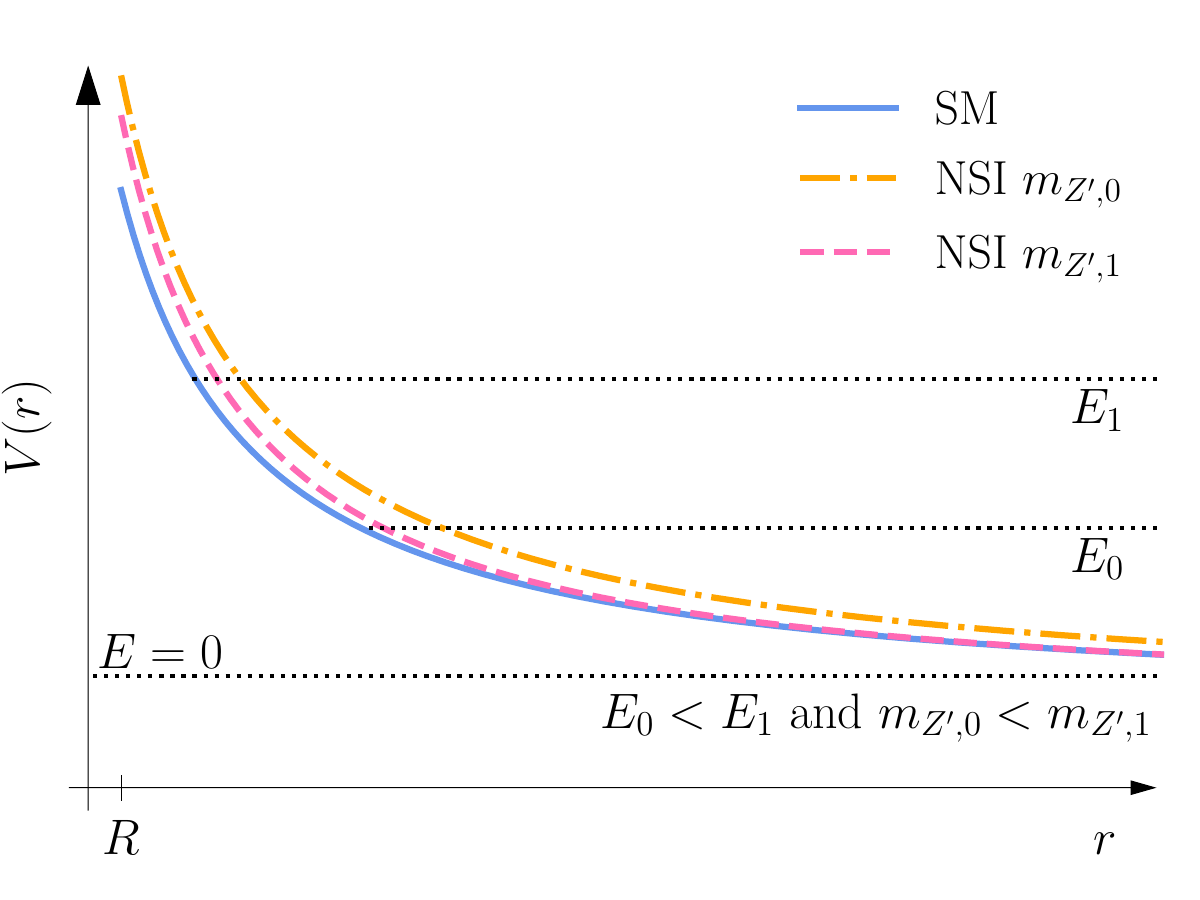}
 \caption{Standard Model Coulomb potential (solid blue line) and the modified potential shown as a function of proton separation. Modified, non-standard interaction (NSI) potentials correspond to a low mass non-standard vector boson (dash-dotted orange line) and a high mass mediator (dashed pink line). The horizontal dashed lines provide examples of the classical turning point for different energies. For reference, the nuclear radius (nuclear touching point) is marked with $R$. 
 }
\label{fig:sketch}
\end{figure}

Figure \ref{fig:sketch} illustrates the radial evolution of the Coulomb potential between two protons in the absence of the NSIs with a solid blue line. This figure also shows the NSI modification of the barrier for a vector mediator with low (high) mass designated with a dash-dotted orange (dashed pink) line, both with the same coupling. The Coulomb potential increases with respect to the Standard Model case regardless of the new vector mediator mass. However, the range over which the growth is affecting the potential depends on the mediator mass; for higher masses, it is shorter than for lower ones. Additionally, the fraction of the barrier that the incoming proton needs to penetrate for the interaction to occur depends on the energy of the incoming proton; higher energies have an advantage over the lower ones. For the scalar mediator, we expect a decrease in the repulsive Coulomb potential, making penetration easier for a given entrance channel center of mass energy.

Any suppression or enhancement in the barrier will have ramifications for the rate of $p+p\rightarrow \mathrm{D} + e^{+} + \nu_{e}$. Given the some three order of magnitude discrepancy between thermal kinetic energies ($\sim {\rm keV}$) of the proton reactants in the Sun and the Coulomb energy at proton-proton contact ($\sim {\rm MeV}$), quantum mechanical tunneling through this barrier leads to exponential suppression of the rate of this reaction. Clearly, even modest NSI modification of this barrier may have significant consequences for solar structure and evolution.

\section{Standard Solar Model and beyond}
\label{sec:Standard Solar Model and beyond}

The fusion of hydrogen nuclei into helium is the primary source of energy in the Sun \cite{Atkinson:1929,Bethe:1939bt,Bethe:1938yy,Weizsacker:1937,Weizsacker:1938,Chandrasekhar:1939,PhysRev.53.608}. It is the only known process that suffices to continuously produce the needed amount of energy for Sun to shine for $4.55 \times 10^{9}$~yrs. The confirmation that this nuclear fusion process occurs inside the Sun came from the observation of solar neutrinos \cite{Davis:1968cp,Hirata:1989zj,Abdurashitov:1994bc,Anselmann:1992um,Arpesella:2008mt,Agostini:2018uly,Agostini:2020mfq} and the discovery of neutrino oscillations \cite{Fukuda:1998mi,Ahmad:2002jz}. 
For a recent review on solar neutrinos, see Ref.~\cite{Vitagliano:2019yzm}.

As outlined above, the primary energy generating reaction sequence is the proton-proton chain ($pp$-chain) and the key first step in this chain is nuclear weak reaction 
\begin{equation}
\label{eq:pp_rection}
    p + p \rightarrow \mathrm{D} + e^{+} + \nu_e \ .
\end{equation} 
Note that $0.24\%$~\cite{Vinyoles:2016djt,Bergstrom:2016cbh} of the time the weak interaction flow will go through the $pep$ process, $p+e^-+p \rightarrow D+\nu_e$, instead of the $pp$ reaction in the above equation. Both reactions have essentially the same barrier penetration physics and will be treated accordingly in what follows.

The rate of this reaction depends on two factors, the weak interaction cross section (and weak nuclear matrix element), and the quantum tunneling probability. 
The overall rate is obtained by integrating over the entrance channel proton energy spectrum,
\begin{equation}
\label{eq:pp_reaction_rate}
    \Gamma_{pp} = \langle \sigma v \rangle = \int_0^{\infty} dE \sigma(E) v(E) u(E) \ , 
\end{equation}
where $\sigma$ is the $pp$ interaction cross section, $v(E)$ is the relative velocity between the particles, and $u(E)$ is the Maxwell-Boltzmann distribution written in terms of the relative energy between two protons in the center of mass frame, $E$, 
\begin{equation}
    u(E) = \frac{2}{\sqrt{\pi}} \frac{\sqrt{E}}{(k_{\rm b} T)^\frac{3}{2}} e^{-\frac{E}{k_{\rm b} T}} \ ,
\end{equation}
where $k_{\rm b}$ is the Boltzmann constant, and $T$ is the temperature of the medium.  
It should be noted that $\sigma$ is the product of the weak cross section and the tunneling probability, both of which are calculated from first principles and cannot be measured. In this paper, we keep the weak interaction cross section (transition matrix element) unchanged because we assume that the NSI is not isospin violating. Instead, we investigate quantum tunneling probability modification stemming from physics beyond the Standard Model. In the following subsection, we discuss the computation of the tunneling probability through the standard Coulomb barrier, closely following Ref.~\cite{Clayton}\footnote{For a recent exposition on the topic see Ref.~\cite{weinberg_2019}.}. We then discuss the techniques we employ to estimate the effect of an NSI contribution to the barrier.

\subsection{Quantum tunneling probability}

We set the stage for NSI effects by giving a quick overview of the standard nuclear reaction rate Coulomb barrier penetration treatment. The wave function describing the relative motion (non-relativistic kinematics) of 2 particles $A$ and $B$ with a relative angular momentum $l$ in spherical coordinates
\begin{equation}
\label{eq:wavefunction}
    \psi (r, \theta, \phi) = \frac{\chi_l (r)}{r} Y_{l}^{m} (\theta, \phi) \,
\end{equation}
for a central field problem can be separated into the angular and radial part, where $Y_l^{m} (\theta, \phi)$ are the spherical harmonics and $\chi_l(r)$ satisfies a wave equation 
\begin{equation}
\label{eq:wave_equation}
    \frac{d^2\chi_l}{dr^2} + \frac{2\mu}{\hbar^2} \left( E - V_l(r) \right) \chi_l(r) = 0 \ .
\end{equation}
The effective radial potential $V_l(r)$ entering the above equation
\begin{equation}
\label{eq:effective radial potential}
    V_l(r) =
    \begin{cases}
        \frac{l(l+1)\hbar^2}{2\mu r^2} + \frac{Z_1 Z_2 e^2}{r} & {r > R} \\
        \frac{l(l+1)\hbar^2}{2\mu r^2} + V_c + V_\mathrm{nuc} & {r < R} \ ,
    \end{cases}
\end{equation}
where $\mu$ is the reduced mass, $Z_i$ charge number for reactant nucleus $i$, $e$ is the elementary electric charge, $\hbar$ Planck constant, $V_c$ is the Coulomb potential, $V_\mathrm{nuc}$ the nuclear potential, and $R$ is the nuclear radius (separation of the nuclear centers at contact). In order for the interaction to occur the incoming, entrance channel, particles need to penetrate the potential barrier.
The penetration factor is defined as,

\begin{equation}
\label{eq:penetration factor}
    P_l = \frac{\chi_l^*(\infty) \chi_l(\infty)}{\chi_l^*(R) \chi_l(R)} \ .
\end{equation}
Note that the penetration factor depends only on the values of the $\chi_l(r)$ ratio for $r > R$, i.e., outside the standard nucleon-nucleon strong potential range, where the wave functions are effectively Coulomb waves. 
However, when the entrance channel kinetic energy is well below the barrier height at contact,  $E \ll V_l(R)$, the penetration factors can be calculated to high precision using the WKB (Wentzel, Kramers, and Brillouin) method. (See Ref.~\cite{Landau:1991wop} for a discussion of approximate differential equation solutions using WKB.)
The solution found using this method yields 
\begin{equation}
\label{eq:WBK_both_sides}
    \chi_l(r) = 
    \begin{cases}
        A \frac{e^{i\pi / 4}}{\sqrt{\left( V_l(r) - E \right)}} \exp {\left( \int_{r}^{R_0} \sqrt{-f(r^\prime)} dr^\prime \right)} & {V_{l}(r) > E} \\
        A \frac{1}{\sqrt{\left(E - V_l(r) \right)}} \exp {\left( i \int_{R_0}^r \sqrt{f(r^\prime)} dr^\prime \right)}& {V_{l}(r) < E} \ ,
    \end{cases}
\end{equation}
where 
\begin{equation}
\label{eq:WBK_f}
    f(r) = \frac{2\mu}{\hbar} \left( E - V_l(r) \right).      
\end{equation}
By inserting eq.~\ref{eq:WBK_both_sides} and eq.~\ref{eq:WBK_f} into eq.~\ref{eq:penetration factor}, the penetration factor becomes 
\begin{equation}
\label{eq:penetration_factor_WBK}
    P_l(r) = \sqrt{ \frac{V_l(r) - E}{E}} \exp{\left(- \frac{2\sqrt{2\mu}}{\hbar} \int_{R}^{R_0} \sqrt{V_l(r) - E}\, dr \right)} \ .
\end{equation}

The quantum tunneling probability for the initial state $l=1$ is suppressed in comparison to the $l=0$ initial state because of the centrifugal barrier in addition to the Coulomb barrier. Consequently, we focus here on s-wave, $l=0$ initial states. By performing the integral in the exponent for $l=0$ and replacing the numerator in the factor before the exponent, $\sqrt{E_c - E}$, by $\sqrt{E_c}$ (as this is in better agreement with the exact solution), where $E_c = {e^2\, Z_1\, Z_2}/{R} \approx 1.44\,{\rm MeV}\, (Z_1\, Z_2)/ (R/{\rm fm})$, the penetration factor is
\begin{equation}
\label{eq:penetration_factor_l_0}
    P_0 \approx \frac{E_c}{E} \exp{\left[-\frac{4 e^2}{\hbar v} \left( \frac{\pi}{2} - \sin^{-1}{\sqrt{\frac{E}{E_c}}} -\sqrt{\frac{E}{E_c}} \left(1- \sqrt{\frac{E}{E_c}} \right) \right) \right]}\approx\frac{E_c}{E}\exp{[-W_0]},
\end{equation}
where the last approximation follows when $E_c \gg E$, as is appropriate in the solar core, and where
\begin{equation}
    W_0 \equiv {\left(\frac{2 \pi e^2}{\hbar v}\right)}\cdot Z_1\cdot Z_2 \equiv b E^{-1/2}.
    \label{W_0}
\end{equation}
Because our focus here is on the $pp$-chain first step reaction, we set $Z_{1}=Z_{2}=1$ in expressions that subsequently appear in this paper.

In the non-relativistic limit, again valid in the solar core, the proton-proton relative entrance channel velocity $v$ can be written as $\sqrt{2E/\mu}$. Local thermodynamic equilibrium in the Sun dictates that the probability distribution for this relative velocity is Maxwellian. In turn, this means that this distribution is an exponentially {\it decreasing} function of the relative center of mass kinetic energy $E =\frac{1}{2} \mu v^2$ of the two protons. This feature, together with a rapid rise in the penetration factor with $E$, guarantees that the cross section (or probability) for the reaction will be proportional to the sharply-peaked function,
\begin{eqnarray}
h(E) \equiv \exp(-E/k_{\rm b}T) \times \exp(-b/\sqrt{E}).
\end{eqnarray}
The peak energy of $h(E)$, the {\it Gamow peak}~\cite{1928ZPhy...51..204G}, is at energy
\begin{eqnarray}
E_{0} = \left(\frac{b\, k_{\rm b} T}{2}\right)^{2/3}.
\end{eqnarray}
From this, it is clear that NSI modifications to the barrier, and hence the reaction rate, will manifest as an altered value of the parameter $b$ in the above. In the Standard Model case this factor is $b = {\left( \sqrt{2}\,\pi\,Z_1\, Z_2\, e^2/(\hbar c) \right)}\, \mu^{1/2} \approx 0.702\,{\rm MeV}^{1/2}$, with the latter approximation pertaining to the proton-proton reaction with $Z_1=Z_2=1$ and reduced mass $\mu$ taken to be half the bare proton rest mass.

\section{Effective dependence on the new coupling}
\label{sec:Effective dependence on the new coupling}

With the new mediator that couples to protons, we can expect an additional term in the Standard Model potential, eq.~\ref{eq:effective radial potential}. This non-standard potential will act as a additional repulsive ($Z^\prime$) or attractive ($\phi$) force: 
\begin{equation}
\label{eq:effective radial potential_new}
V_l(r) =
\begin{cases}
\frac{l(l+1)\hbar^2}{2\mu r^2} + \frac{e^2}{r} \pm \frac{g^2}{r} e^{-m_{\{Z^\prime, \phi\}} r} & {r > R} \\
\frac{l(l+1)\hbar^2}{2\mu r^2} + V_c + V_\mathrm{nuc} + V_\mathrm{NSI} & {r < R} \ .
\end{cases}
\end{equation}
The biggest change to the $pp$ reaction rate induced by the new mediator will arise in the penetration factor eq.~\ref{eq:penetration_factor_l_0}. 

In the limit of low mediator mass and small coupling $g$, the penetration factor does not depend on the mediator mass and the new, modified, penetration factor has a simple analytical form expressed as 
\begin{equation}
W_0^\mathrm{NSI} = W_0 \cdot \left(1 \pm \frac{g^2}{e^2}\right) \ ,
\label{mod}
\end{equation}
where \lq\lq $+$\rq\rq\ and \lq\lq $-$\rq\rq\ in this expression correspond to repulsive ($Z^\prime$) and attractive ($\phi$) potential modifications, respectively. As the mediator mass increases, the range of the NSI addition to the potential shrinks and the integral in the penetration factor cannot be rendered into the simple analytical form in eq.~\ref{mod}.
However, it is possible to compute the integral numerically for a given energy and then compare the value of the integral with and without the NSIs. 
For a given energy and temperature, this gives a good estimate of the change in the reaction rate arising from the NSIs. 
For each energy, we compare the obtained value with the one calculated without the new mediator $(g = 0)$ and define 
\begin{equation}
\label{eq:alpha}
    \alpha \equiv \frac{\left| W_{0,\mathrm{NSI}}^{\frac{2}{3}} - W_{0,\mathrm{SM}}^{\frac{2}{3}} \right| }{W_{0,\mathrm{SM}}^{\frac{2}{3}}} \ ,
\end{equation}
where NSI and SM denote calculations of the full penetration factor with and without, respectively, the mediator modification to the overall potential.

To summarize the logic used in this paper: The $pp$-chain reaction rate depends on the quantum tunneling probability, which can be adjusted in the numerical simulations of evolution of the Sun. For each value of the quantum tunneling probability that is different from the Standard Model value, it is possible to calculate the loci of points in the $(g, m_{\{Z^\prime, \phi\}})$ plane that give a specific change in quantum tunneling probability. 
For example, the decrease in the quantum tunneling probability caused by a vector mediator results in a hydrostatic equilibrium in the Sun at higher temperatures, while the opposite is true in the case of scalar mediators.

Apart from the change in the temperature in the core of the Sun, which can be {indirectly} measured using the measured solar neutrino fluxes, as described in the next section, the change in temperature can also affect the rate of other reactions that contribute to hydrogen fusion. In the core of the Sun, the CNO cycle contributes at a sub-percent level to hydrogen fusion energy generation and neutrino production. In stars that are more than 1.3 times the mass of the Sun, the CNO cycle dominates over the $pp$-chain in energy generation via fusion. Where the bottleneck in the $pp$-chain is the weak interaction first step, the bottleneck(s) in the CNO cycle are the Coulomb barriers in the electromagnetic $(p, \gamma)$ and strong $(p, \alpha)$ nuclear reactions in the cycle. These involve C and N with $Z=6$ and $7$, respectively, making for larger Coulomb barriers and, hence, high temperature sensitivity in the associated penetration factors and reaction rates. Of course, an NSI mediator also would modify these CNO nuclear reaction rates.
{ However, these modifications should be already included in the cross sections for those reactions, as those cross sections are extrapolated from nuclear physics measurements~{\cite{Angulo:2001zrh,PhysRevC.67.065804,Lemut:2006va,Imbriani:2005jz,Runkle:2004mx,Adelberger:2010qa,Azuma:2010zz,Marta:2011hq,Li:2016jzu}} (see Section~\ref{sec:intro}). Nevertheless, discussion in Section~\ref{sec:Results} we re-simulated our results for a case with modified Coulomb barriers not only in the $pp$ and $pep$ reactions but also in the $^{14}\mathrm{N} (p, \gamma) ^{15}\mathrm{O}$ reaction}.

{ The strong temperature sensitivity of the rates of the key processes in the CNO cycle compared to the key first step in the $pp$-chain is what we use to constrain the NSIs.}
This sensitivity is not altered by the NSIs. Likewise for the high energy ${^8{\rm B}}$ neutrinos from the side branch of the $pp$-chain that arises from the Coulomb barrier penetration-sensitive, and hence temperature sensitive, $^7{\rm Be}(p, \gamma){^8{\rm B}}$ reaction.

Finally, it should be noted that the nuclear reaction rates that play a role in the CNO cycle cannot be calculated with sufficient accuracy from first principles and are measured in the laboratory. { As mentioned before,} any effect of the NSI would already be incorporated in the present rate calculation, or be so small at the level considered here that its effect on these CNO reactions would be within the current errors in the extrapolation of their S-factors to the range of most effective energies in the Sun. In our analysis we neglect NSI effects on the HEP reaction, ${^3{\rm He}} + p \rightarrow {^4{\rm He}} + e^++\nu_e$, contribution to the solar neutrino flux \cite{2004ARNPS..54...19K,1998PhLB..436..243B,Agostini:2018uly}.

The CNO cycle positron decay-produced neutrinos were measured in the Borexino experiment \cite{Agostini:2020mfq}. These CNO cycle positron decays are $^{13}{\rm N}\rightarrow {^{13}{\rm C}}+e^++\nu_e$ and $^{15}{\rm O}\rightarrow {^{15}{\rm N}}+e^++\nu_e$. Although the error-bars on the CNO neutrinos are large, we can still use the data to constrain the parameters of the NSIs. In the next section, we present the constraints on these NSI parameters using the temperature in the core of the Sun as well as the ratio of CNO to $pp$-reactions.

\section{Results}
\label{sec:Results}

We consider the effects of a non-standard vector and scalar boson mediator in turn. Let us first discuss the general effects of the vector mediator case. The vector mediator leads to an additional repulsive force in the effective proton-proton interaction. 

The additional repulsive potential implies that the interior of the Sun must contract to a higher temperature before an equilibrium state is reached where nuclear energy generation is balanced by energy transport. The increased temperature has two distinct observable effects. We exploit these effects to explore and even constrain the effects of a non-standard addition to the repulsive force.

We simulated the evolution of a solar mass star using the {\it Modules for Experiments in Stellar Astrophysics} (MESA) code \cite{Paxton2011,Paxton2013,Paxton2015,Paxton2018,Paxton2019}.

In this code, we changed only the $pp$ and $pep$ reactions' rate exponent factors to include the effects of the non-standard mediators. In the Standard Model case, those interaction rates scale as \cite{Clayton}
\begin{equation}
\label{eq:exponent_rate}
\Gamma_{pp} \propto \exp\left(-3.381 (1 + \Delta) \; \left(\frac{T}{10^9 \; \mathrm{K}}\right)^{\frac{1}{3}}\right) \ ,
\end{equation}
with $\Delta=0$. The new mediator physics can be incorporated by taking $\Delta > 0$ for the vector boson mediator and $\Delta < 0$ for the scalar mediator. By assuming that the energy dependence of the calculated $W_0^\mathrm{NSI}$ will not change significantly, we obtain the relation $\Delta \approx \alpha$, with $\alpha$ defined in eq.~\ref{eq:alpha}.

We are working under the assumption that the key weak interactions, $pp$ and $pep$, have the largest impact on the outcome of the nuclear fusion inside the core of the Sun and that they have not been measured in the laboratory at stellar energies. We leave further studies of the impact of such NSIs on stellar evolution in general for future work. In MESA, we did not change any of the default values of the initial parameters except for the initial metallicity, which in our study we vary over the range $Z_\mathrm{init}= 0.014-0.02$, a range consistent with the theoretical uncertainty \cite{Vinyoles:2016djt} and the inferred observational uncertainty in this quantity \cite{Agostini:2018uly}.
We evolve a $1\,{\rm M}_\odot$ stellar model for a time corresponding to the known solar age, $4.55 \times 10^9~\mathrm{yrs}$, and then compare the structure and parameters of this configuration to solar measurements.

\begin{figure}[t]
 \centering
 \includegraphics[width=\columnwidth]{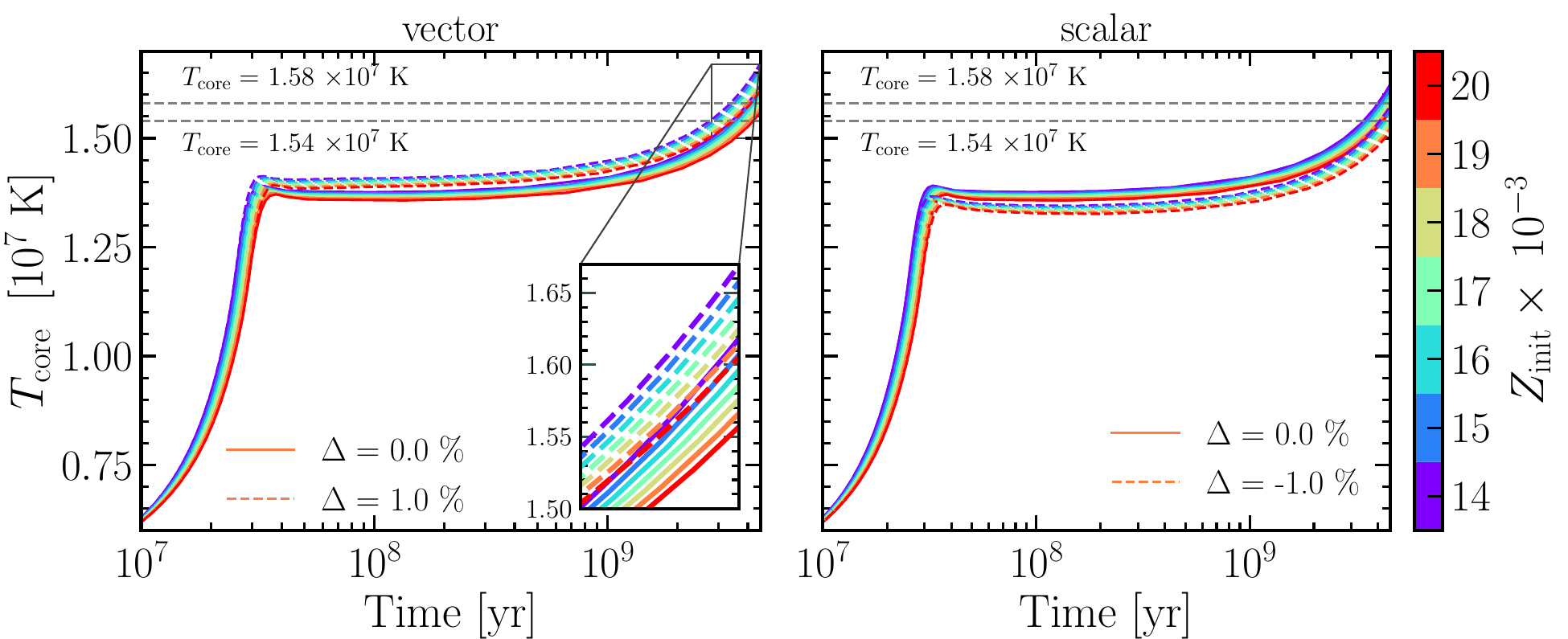}
 \caption{The temporal evolution of a 1~$M_\odot$ star's core temperature, $T_{\rm core}$, in the presence of the new vector (scalar) mediator is shown in the left (right) panel. In both panels the standard model case corresponds to the solid lines ($\Delta = 0$) and the NSI case ($\Delta \neq 0$) results are presented as dashed lines. Initial metallicity, $Z_\mathrm{init}$, color scale is given at far right. The colors reflect the various initial values for metallicity, taken in the range $Z_\mathrm{init} = 0.014 - 0.020$. The grey dashed horizontal lines mark the values consistent with the current solar core temperature. The positive (negative) change in the barrier height, corresponding to $\Delta > 0$ ($\Delta < 0$), causes the core to heat up, i.e., reach an equilibrium state at a higher (lower) temperature.}
\label{fig:temperature_radial}
\end{figure}

The left panel of Figure~\ref{fig:temperature_radial} shows the temporal evolution of the Sun's core temperature ($T_\mathrm{core}$) for the changed exponent factors inside the penetration factor, eq.~\ref{eq:penetration_factor_l_0}, resulting from the presence of a new vector boson mediator ($\Delta > 0$). As the barrier increases, the core temperature follows that trend. That behavior is a consequence of the need for the higher proton energies required to penetrate the modified Coulomb barrier. This figure also demonstrates that for smaller initial metallicities (different colors), the core's temperature increases regardless of the change in the exponent. We note that small changes in the barrier could be compensated with an altered initial metallicity. For example, a higher barrier could be interpreted as a lower metallicity and no change in the Coulomb potential. 

On the contrary, for the scalar mediator (right panel of Figure~\ref{fig:temperature_radial}, $\Delta < 0$), and a consequent decrease in the Coulomb barrier, the temperature of the core of the Sun does not grow to as high a value as without the presence of the non-standard mediator. This is because efficient and sustainable fusion is possible with lower proton energies. Analogously to the vector case, we see that lower initial metallicities result in higher core temperatures, even without an NSI alteration of the penetration factor.

The temperature at the core of the Sun can be estimated by measuring the neutrino fluxes originating from the side branches of the $pp$-chain involving ${}^{7}\mathrm{Be}$ and ${}^{8}\mathrm{B}$ \cite{Bahcall:1996vj,TurckChieze:2010gc}. The nearly mono-energetic beryllium neutrino flux is a result of the electron capture reaction,
\begin{eqnarray}
\label{eq:Be_reaction}
{}^{7}\mathrm{Be} + e^{-} \rightarrow {}^{7}\mathrm{Li} + \nu_{e}\ .
\end{eqnarray}
The rate of this reaction depends on the $11^\mathrm{th}$ power of temperature. The Boron-decay neutrino flux resulting from the reaction sequence
\begin{eqnarray}
\label{eq:B8_reaction}
{}^{7}\mathrm{Be} + p \rightarrow {}^{8}\mathrm{B} + \gamma \rightarrow {}^{8}\mathrm{Be}^* + e^{+} + \nu_{e}
\end{eqnarray}
depends on $21^\mathrm{st}-25^\mathrm{th}$ power of temperature. We leverage the high temperature sensitivity of these reaction rates, coupled with the measured neutrino fluxes to probe and constrain our non-standard mediators. We note that the small effects of our mediators on altering the effective Coulomb barriers in these side-branch reactions will not appreciably change their temperature sensitivity, the key effect our probe depends on.   

We estimate the allowed change in the temperature of the core of the Sun with the uncertainties in the theoretical predictions of the side branch neutrino fluxes. The uncertainty in the ${}^8$B neutrino flux is $12\%$, while it is $6\%$ for the ${}^7$Be \cite{Vinyoles:2016djt} flux. Taking into account those values and the steep temperature dependence of the reaction rates, the uncertainty in the temperature of the core of the Sun should be between $0.5-0.6\%$. In the following, we will assume conservatively that the solar core's temperature uncertainty is $1\%$.

\begin{figure}[t]
 \centering
 \includegraphics[width=\columnwidth]{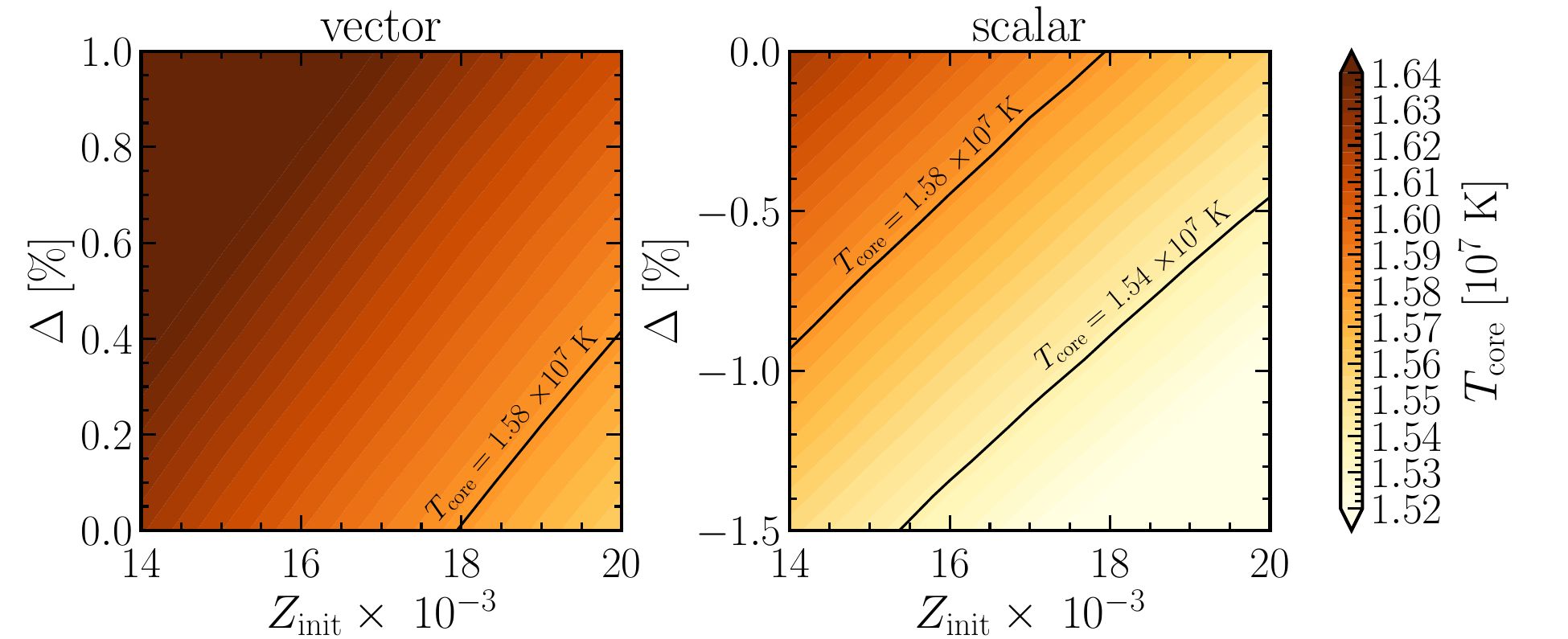}
 \caption{Contours of the current Sun's core temperature, $T_{\rm core}$, shown in the initial metallicity, $Z_\mathrm{init}$, and rate exponent alteration, $\Delta$, plane. Contour color key at far right. Left panel shows the vector mediator case; right panel shows the scalar mediator case. The acceptable (non-ruled out) change in the core's temperature lays under the solid black line in the left-hand panel. The allowed (non-ruled out) region lays between the two solid lines in the right-hand panel, with the upper contour representing the allowed limit on temperature increase, and the bottom one the allowed limit on temperature decrease in this case. 
 }
\label{fig:2D_temperature}
\end{figure}

\begin{figure}[t]
 \centering
 \includegraphics[width=\columnwidth]{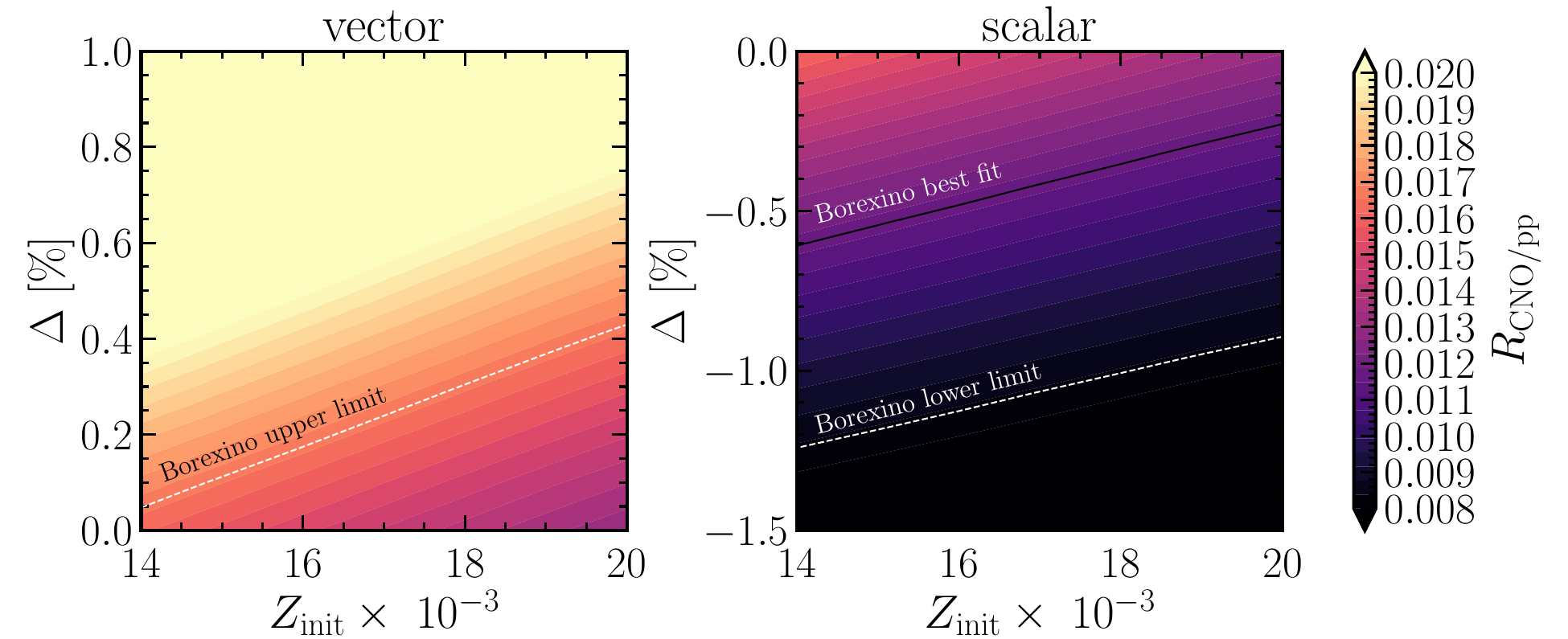}
 \caption{Contours of the current Sun's CNO to $pp$ energy generation ratio, $R_\mathrm{CNO/pp}$, shown in the initial metallicity $Z_\mathrm{init}$ and $\Delta$ plane. Contour color key at far right. Left panel shows the vector mediator case; right panel shows the scalar mediator case. The left panel shows the Borexino upper limit (white dashed line). The right panel shows the Borexino best fit (dark line) and lower limit (white dashed line). The ratio $R_\mathrm{CNO/pp}$ is larger for the low metallicities than for the high ones, as a result of the corresponding bigger change of the core temperature for the lower initial metallicity cases.}
\label{fig:2D_CNO_PP}
\end{figure}  

Figure~\ref{fig:2D_temperature} shows contours of temperature for the stellar (solar) core in the initial metallicity, $Z_\mathrm{init}$, rate exponent alteration, $\Delta$, plane. This figure shows the effects of a new vector mediator in the left panel, and the scalar mediator effects in the right panel. These results are calculated with a consistent stellar model evolved to the current age of the Sun. As illustrated in Figure~\ref{fig:temperature_radial}, the temperature of the Sun's core increases (decreases) in the presence of the non-standard vector boson (scalar) mediator. The acceptable change in the core's temperature lays under the solid black line representing the allowed increase of the solar core's temperature for the non-standard vector mediator. In the case of the scalar mediator, the allowed (non-ruled out) region lays between the two solid lines, where the upper contour represents the allowed limit on temperature increase, and the bottom one the allowed limit on temperature decrease. 

The Standard Solar Model predicts the flux of electron neutrinos from the $pp$ reaction to be between $5.98 - 6.03 \times 10^{10} \; \mathrm{cm}^{-2} \; \mathrm{s}^{-1}$ depending on the initial metallicity \cite{Vinyoles:2016djt}. The flux measured by the Borexino collaboration is in agreement with those values \cite{Agostini:2018uly}. In addition, the CNO neutrino flux has also recently been observed, $7_{-2}^{+3} \times 10^{8} \; \mathrm{cm}^{-2} \; \mathrm{s}^{-1}$ \cite{Agostini:2020mfq}. Those observations permit us to calculate the observational constraint on the CNO to $pp$ neutrino flux ratio, $R_\mathrm{CNO/pp}$. We find the allowed region of the ratio to be in a range $0.0083 - 0.0167$, with the $68\%$~C.L. best fit value being $0.0117$ \footnote{This calculation assumes that the survival probabilities for $pp$ and CNO neutrinos do not vary significantly, and taking that effect into account is subleading to the uncertainty of the detected CNO flux.}.

We investigated the impact of the non-standard mediators on the energy ratio between the CNO to $pp$ ($R_\mathrm{CNO/pp}$) reactions at the current solar age. Our findings are depicted in Figure~\ref{fig:2D_CNO_PP}. Similar to Figure~\ref{fig:2D_temperature}, the presence of the vector boson (scalar) mediator increases (decreases) the probed parameter. For the positive change in the Coulomb barrier height, the parameter $R_\mathrm{CNO/pp}$ increases due to the higher temperature of the core, which allows the CNO reactions producing neutrinos to proceed more rapidly than in case of a lower core temperature. On the other hand, the presence of a non-standard scalar mediator suppresses the efficiency of CNO neutrino production because of the lower temperature of the solar core. Note that lower metallicities are characterized by a higher $R_\mathrm{CNO/pp}$ regardless of the $\Delta$ value for both kinds of the mediators. This might be counter-intuitive at first since for lower metallicities, we would expect smaller $R_\mathrm{CNO/pp}$. However, the CNO cycle requires high temperatures to occur efficiently \cite{Weizsacker:1937,Weizsacker:1938,Bethe:1939bt}, and that is the case for the low metallicities as can be seen in Figure~\ref{fig:2D_temperature}.

\begin{figure}[t]
 \centering
 \includegraphics[width=\columnwidth]{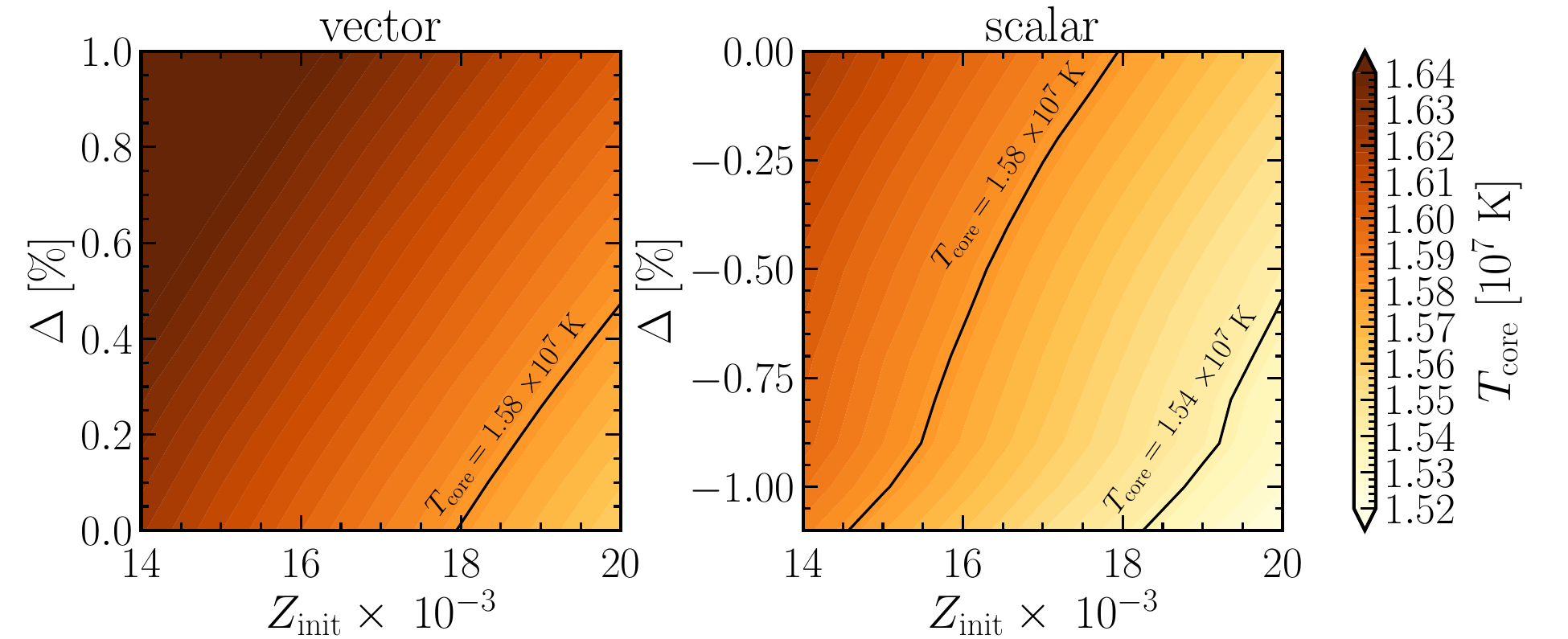}
 \caption{Contours of the current Sun's core temperature, $T_{\rm core}$, shown in the initial metallicity, $Z_\mathrm{init}$, and rate exponent alteration in the deuteron fusion reaction, $\Delta$, plane; results for calculations where the change in the CNO  bottleneck reaction, $^{14}\mathrm{N} (\mathrm{p}, \gamma) ^{15}\mathrm{O}$ was also included. Contour color key at far right. Left panel shows the vector mediator case; right panel shows the scalar mediator case. The acceptable (non-ruled out) change in the core's temperature lays under the solid black line in the left-hand panel. The allowed (non-ruled out) region lays between the two solid lines in the right-hand panel. Changes with respect to the default case (rate exponent alterations only in the $pp$ and $pep$ reactions)
are not significant.}
\label{fig:2D_temperature_CNO}
\end{figure}

\begin{figure}[t]
 \centering
 \includegraphics[width=\columnwidth]{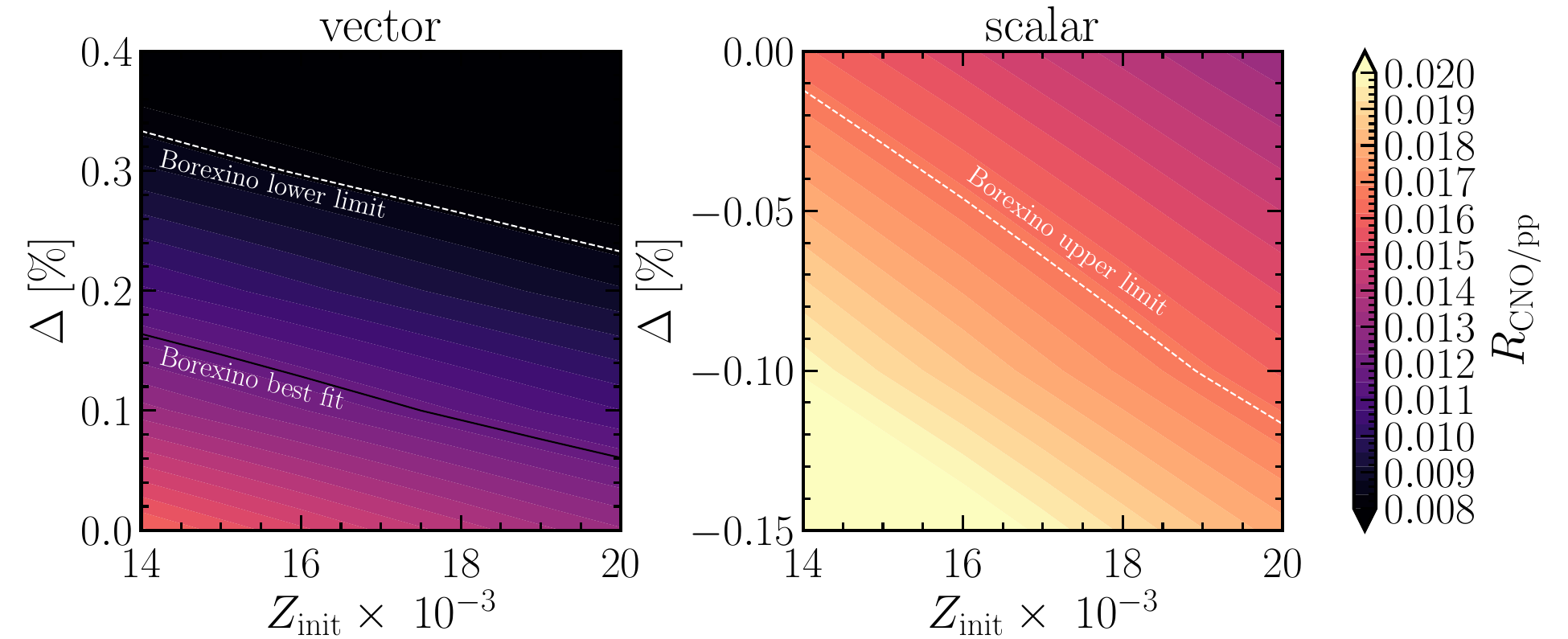}
 \caption{Contours of the current Sun's CNO to $pp$ energy generation  ratio, $R_\mathrm{CNO/pp}$, shown in the initial metallicity $Z_\mathrm{init}$ and $\Delta$ plane for calculations where the change in the CNO bottleneck reaction, $^{14}\mathrm{N} (\mathrm{p}, \gamma) ^{15}\mathrm{O}$ was also included. Contour color key at far right. Left panel shows the vector mediator case; right panel shows the scalar mediator case. The left panel shows the Borexino lower limit (white dashed line) and the Borexino best fit (dark line). The lower limit (white dashed line) is on the right panel. The ratio $R_\mathrm{CNO/pp}$ is larger for the scalar case than for the vector case, as a result of the corresponding decrease in the Coulomb barrier.}
\label{fig:2D_CNO_PP_CNO}
\end{figure}

Additionally, for completeness, we have re-simulated our models to obtain the results for changed Coulomb barrier heights, not only in the $pp$ and $pep$ reactions but also in the bottleneck CNO reaction - $^{14}\mathrm{N} (\mathrm{p}, \gamma) ^{15}\mathrm{O}$.
{This is done to investigate the sensitivity of the obtained results to the experimental and extrapolation uncertainties in the total S-factor for the $^{14}\mathrm{N} (\mathrm{p}, \gamma) ^{15}\mathrm{O}$ reaction. With the non-standard interactions invoked here actually present, the measured laboratory cross sections would already have their effect included. But we must add the caveat that the non-standard physics may introduce a different energy dependence not taken into account in the standard extrapolation procedures. This merits further investigation.}

The simplified value of the $b$ parameter for {$^{14}\mathrm{N} (\mathrm{p}, \gamma) ^{15}\mathrm{O}$} reaction is $b_\mathrm{CNO} \simeq 6.72~\mathrm{MeV}^{1/2}$. For each $\Delta$ change in the Coulomb barrier of the $pp$ and $pep$ reactions ($b = b_{pp} = 0.702~\mathrm{MeV}^{1/2}$) the change in the $^{14}\mathrm{N} (\mathrm{p}, \gamma) ^{15}\mathrm{O}$ is $(b_\mathrm{CNO} / b_\mathrm{pp})^{2/3} \simeq 4.5$ times higher.
A $1\%$ shift in the $pp$ and $pep$ barriers corresponds to a $4.5\%$ change in the CNO barrier height, 
 which corresponds to $14.5\%$ change in the reaction rate for $T = 1.56 \times 10^7~\mathrm{K}$.
For higher temperatures (energies), the change decreases. 

The implemented shifts in the Coulomb barrier coming from the non-standard mediators introduce cross section modifications that are {comparable to the experimental uncertainty, $\mathcal{O}(10\%)$~\cite{Angulo:2001zrh,PhysRevC.67.065804,Lemut:2006va,Imbriani:2005jz,Runkle:2004mx,Adelberger:2010qa,Azuma:2010zz,Marta:2011hq,Li:2016jzu}\footnote{Ref.~\cite{Frentz:2020pce}, which presents new measurements of the lifetimes of the excited states in $^{15}$O, also promises a forthcoming paper with an updated estimate of the total astrophysical S-factor for the $^{14}\mathrm{N} (\mathrm{p}, \gamma) ^{15}\mathrm{O}$ interaction.}.}
Besides, the cross section extrapolated from the laboratory measurements should already include the effects imposed by the non-standard mediators.

Figures~\ref{fig:2D_temperature_CNO} and \ref{fig:2D_CNO_PP_CNO} illustrate the changes in  the solar core temperature and the CNO to $pp$ energy generation ratio for the re-simulated models.
One can notice that while the changes in the temperature of the solar core are modest with respect to the default case (i.e., with only the barriers in the $pp$ and $pep$ reactions changed), the  behavior of the CNO to $pp$ energy generation  ratio flips. The reason behind this is a more prominent lowering (increasing) of the Coulomb barrier than in the $pp$ and $pep$ reactions.

To calculate the respective bounds on the mass and the coupling of the non-standard mediators, we use the maximal value of allowed $\Delta$ from the Figure~\ref{fig:2D_CNO_PP}. In the case of the vector boson this is $\Delta \simeq 0.4\%$, while for the scalar case it is $\Delta \simeq -1.25\%$. This allows us to rule out the non-standard mediators with masses and couplings that lead to change in the absolute value of $\Delta$ larger than $0.4\%$ (1.25\%) for the vector boson (scalar) mediator. 

\begin{figure}[t]
 \centering
 \includegraphics[width=\columnwidth]{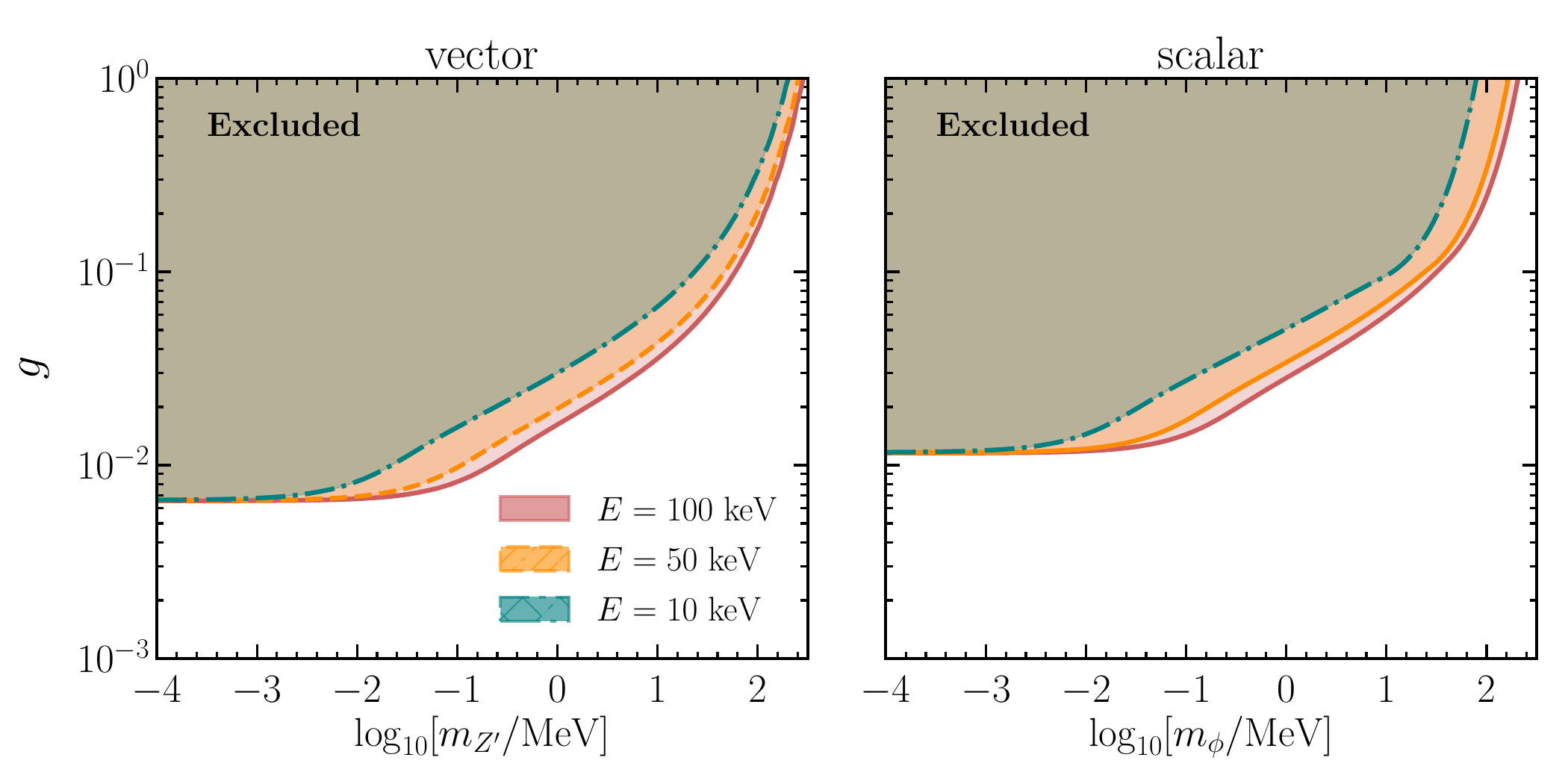}
 \caption{Bounds (colored regions) on the new vector boson (left panel) and scalar (right panel) coupling to protons shown in the mediator mass, $m_{\{ Z^\prime, \phi\}}$, coupling plane, $g$, plane. The bounds are calculated for three different proton energies: 100~keV (solid red), 50~keV (orange dashed), and 10~keV (dash-dotted green). The region of the parameter space excluded in the vector boson case is bigger because of the tighter constraints on the temperature and $R_\mathrm{CNO/pp}$ for that case.}
\label{fig:Bounds}
\end{figure}

The respective limits on the non-standard vector and scalar mediators are depicted in the left and right panels of Figure~\ref{fig:Bounds}. The bounds were calculated for three different proton energies: 10~keV (green, dash-dotted line); 50~keV (dashed, orange line); and 100~keV (solid, red line). These energies are representative of the conditions extant in the solar core. In this figure, the excluded parameter regions are shaded. In both cases, the shape of the ruled-out region follows the same trend. In the low mediator mass limit, the eliminated region depends only on the coupling to the new mediator. This can be understood by examining eq.~\ref{nsipot}. In the limit of $m_{\{Z^\prime, \phi\}} \ll r$, this equation shows that the potential depends only on the coupling to the new particle, as the exponent approaches unity. With increasing mass for the mediator, the range of the new force decreases, hence larger couplings are needed to make a measurable change in the final penetration factor. The figure also shows that the excluded region shrinks as the mediator mass increases for smaller proton energies used in the calculation. That feature is a result of a smaller region of the potential affected by the non-standard mediator in the intermediate and high mass limit.

Depending on the nature of the NSIs, the additional force could result in other observable effects. However, the additional force between two protons is the most generic outcome that cannot be avoided in any model of physics beyond the Standard Model that involves a mediator that can couple with nucleons. In the next section, we discuss other observable quantities that may be affected by the NSIs considered in this paper.

\section{Discussion}
\label{sec:Discussion}

We have considered the case where new NSIs alter the effective interaction between nucleons, in particular in the proton-proton interaction channel with our attention on how that new interaction alters penetration through the Coulomb barrier. The arguments used in this work to constrain NSIs in the proton-proton channel are largely independent of model features. However, general nuclear physics considerations dictate some qualifications on possible NSIs in nucleon-nucleon interactions. 
This includes isospin violation of sufficient magnitude that it is not allowed by other constraints.
Proposed NSIs obviously cannot violate anything measured or otherwise known about nuclei.

As discussed above for our NSIs, the force between two protons is repulsive if mediated by a vector boson, and attractive if mediated by a scalar. In fact, the force mediated by a scalar is always attractive, but the force mediated by a vector boson is repulsive only between particles of {\it like charge}. In this paper, we have not discussed or specified whether protons and neutrons have the same or opposite charge under the new coupling to a vector boson; in principle either is possible. Any feature of NSIs that break isospin symmetry are problematic at some level.  

For example, depending on the details of the model, the introduction of a new force in the neutron-proton channel could change the binding energy or quadrupole moment of the deuteron. It should be noted, however, that the additional couplings considered here, and not ruled out by the considerations in this paper, are very small. Effective field theory and other {\it ab initio} structure calculations, as described above, could model these effects. However, given experimental errors, it is not obvious whether these small effects can be constrained. 

Isospin symmetry, inherent in the strong interaction, is broken by the Coulomb interaction in nuclei. 
Mirror nuclei are nuclei with two different atomic numbers, but the same number of nucleons. It is well known that in such nuclei, the Coulomb repulsion between protons leads to a lower binding energy in the case of nuclei with larger atomic number. The energy shift of the ground states and other energy levels in these nuclei associated with the Coulomb repulsion can be calculated using semi-empirical methods. It may be possible to eliminate the possibility of a sub-class of isopsin symmetry violating non-standard interactions on the basis of trends in the energy level and binding energy systematics of nuclei. Parity violation in the NSI nucleon-nucleon channel is likewise problematic and again might be constrained through nuclear structure systematics. 

We, once more, want to bring to the reader's attention that we model the Sun by relying on the publicly available, open-source MESA code~\cite{Paxton2011,Paxton2013,Paxton2015,Paxton2018,Paxton2019}. 
While this allows us to obtain robust qualitative {\it trends} of the NSI effects on the solar evolution, the quantitative results might vary slightly if simulations employ a different solar model.

It would be interesting to see the variation of the trends and bounds derived in our work if a more sophisticated, higher precision solar model code were employed. Examples of more sophisticated codes include those employed in state-of-the-art standard solar models~\cite{Vinyoles:2016djt,Villante:2021ubp}. 
These models also include calibration procedures that rely on adjusting free parameters -- mixing length, initial helium, and metal abundances -- in the models to match the present-day solar observables, i.e, solar luminosity, radius, and surface metal abundances. Our simulations do not include such a procedure. 
The potential impact of the re-calibration on the obtained bounds is left to future work.

Finally, although we focus on the Sun in this paper, it may be possible to get better constraints based on other astrophysical systems that generate energy by hydrogen fusion. This might be achieved, for example, either by studying individual stars or astrophysical objects in general or by statistical analysis of a class of astrophysical objects. One particularly interesting consequence of non-standard interactions between protons could be the change in the hydrogen burning limit in brown dwarfs~\cite{Hayashi1963,kumar}. A change in the magnitude of the repulsive force between protons would change the minimum mass required for the onset of hydrogen burning in low mass stars, which is approximately 0.07~$M_{\odot}$ in the absence of any non-standard interactions.
The theoretical limiting mass separating the objects which can sustain hydrogen burning from those which cannot, i.e., the mass boundary between red dwarfs and brown dwarfs, is close to $0.07~\mathrm{M}_{\odot}$ \cite{Hayashi1963,kumar,RevModPhys.65.301,1993ApJ...406..158B,1993ApJ...403..303L,1995ApJ...446L..35B,Burrows_1997,Chabrier:2000ns}. This limit should be sensitive to the non-standard mediators proposed in our work. However, as a consequence of the large uncertainties in the determination of the red and brown dwarfs masses \cite{Beichman2013,Brandt2020}, the bounds calculated in this work should be more constraining. Additionally, this limiting mass is also sensitive to the initial composition of the star such that metal-rich stars need smaller mass to fuse hydrogen. Another factor complicating using the hydrogen-burning mass limit to constrain the non-standard mediators is the possibility of the existence of the over-massive brown dwarfs \cite{Forbes2019}.

\subsection{Additional constraints}
\label{sec:Additional constraints}

Obviously there are other ways to constrain some of the mediator mass and coupling parameter space probed by our considerations. Here we discuss these alternative constraints that might apply depending on the details of the assumed model.

The limits on the mass and coupling of new leptophobic dark gauge bosons have been studied in Ref~\cite{Rrapaj:2015wgs}. There it was found that mediators with masses below $200~\mathrm{MeV}$ and couplings $ 3 \times 10^{-10} \lesssim g \lesssim 4 \times 10^{-8}$ can be excluded by applying to core collapse supernovae an energy-loss argument \cite{Raffelt:1996wa} and trapping criterion. Those limits test a mediator mass range similar to that considered in this work but for a parameter range with smaller couplings.
Similarly, in Ref.~\cite{Grifols:1988fv}, by assuming a cooling effect in the Sun, it was argued that for mediator mass $m \lesssim 2~\mathrm{keV}$ couplings between $3 \times 10^{-11}  \lesssim g \lesssim 10^{-3}$ can be excluded.

Additional constraints coming from the limits on the branching ratio for pion decays into a photon and a non-standard vector boson should apply in the models with an equal coupling of the new mediator to the first generation of quarks \cite{Dobroliubov:1990ye,Dobroliubov:1987cb}. Those limits~\cite{Gninenko:1998pm,Tulin:2014tya,Farzan:2016wym} are comparable to the ones presented in our work for mediators with masses below $120~\mathrm{MeV}$, in the low mass limit the difference is less than a factor of {six}. However, we used a conservative estimate of the uncertainty on the core temperature of the Sun. For a more relaxed estimation involving $2.5\%$ uncertainty on the flux of neutrinos from the $^8\mathrm{B}$ decay~{\cite{Abe:2016nxk}}, our limits improve and are { worse} by a factor of {$\sim2$} than the ones obtained with pion decays. Moreover, with an improved determination of solar metallicity, and a more precise measuremnet of the CNO flux it is very likely that our limits will improve even more.

In models with equal coupling to both nucleons, the limits on the non-standard mediator coming from the neutron scattering on lead \cite{Schmiedmayer:1988bm,Barbieri:1975xy,Leeb:1992qf} dominate our bounds for {$m\lesssim70~\mathrm{MeV}$}.

In the low mediator mass ranges there are more constraining bounds coming from fifth force searches. The explorations focused on testing the gravity at short distances (see \cite{Lee:2020zjt} and references therein) exclude the region of the parameter space with mediator masses smaller than $\lesssim 5 \times 10^{-3}~\mathrm{eV}$. Additionally, the bounds from the exotic systems such as antiprotonic helium ($\bar{\mathrm{p}} {}^{4}\mathrm{He}^+$) and $\mathrm{HD}^+$ were calculated by comparing the observed and theoretically derived high angular momentum nuclear \lq\lq atomic\rq\rq\ transition frequencies \cite{Salumbides:2013aga,Salumbides:2013dua}. The limits resulting from those estimations can exclude the region of parameter space with $m\lesssim50~\mathrm{keV}$.

We also note that if the new mediator kinetically mixes with the Standard Model $Z$ boson or the electromagnetic field tensor, additional constraints might apply (see, for example, Refs.~\cite{Davoudiasl:2012ag, Abdullah:2018ykz}).

\section{Conclusions}
\label{sec:Conclusions}

There are several conclusions to be drawn from this study. We have leveraged the high sensitivity of quantum tunneling through a barrier, in the limit where the reactant energies are small compared to the barrier height, to connect speculative ideas from BSM particle physics to well measured properties of the Sun. In broad brush, our analysis shows that introducing new vector or scalar mediators in the proton-proton interaction channel can be problematic with respect to the measured solar properties, allowing constraints on the mediator masses and couplings. Of course, there are many other ways to constrain much of the parameter space of mediator character, mass, and coupling probed here. However, each of these modalities for constraint are different, each probing a somewhat different aspect of the underlying physics and each based on differing underlying assumptions.

Of all the nuclear reactions of consequence in the Sun, the first stage of the $pp$-chain reaction is the only one that has not been measured in the laboratory. The possibility of non-standard interaction between two protons cannot be ruled out, and the existence of such non-standard interaction can result in a change of temperature at the core of the Sun. The temperature at the core of the Sun can be { indirectly} measured to within an accuracy {of $0.09\%$ and the theoretical uncertainty of $0.5\%$ from the Standard Solar Models}, and this can be used to put a limit on the parameters of non-standard interactions. In this paper, we error on the side of caution and assume an uncertainty of 1\% in the { determination} of the temperature at the core of the Sun. We would like to emphasize that the goal of this paper is not to reproduce the SSM results but to { highlight the effects of the non-standard mediators on the solar evolution and} demonstrate the utility of the solar measurements in constraining physics beyond the Standard Model. 

In the case of non-standard interaction between protons that is mediated by vector bosons, there is a repulsive force between in the protons in addition to the Coulomb force, while the converse is true in the case of scalar interactions. A higher temperature is required to achieve hydrodynamic equilibrium in the case of a vector mediator and a lower temperature is required in the case of a scalar mediator.
The neutrino flux from the Sun is highly sensitive to the temperature at the core of the Sun. In addition, the ratio of the prevalence of the CNO cycle to $pp$-chain reaction is also sensitive to the temperature.
The conversion of two protons to a deuterium nucleus occurs via weak interaction, but it is highly suppressed because the reaction requires overcoming the Coulomb barrier by quantum tunneling. The addition of non-standard interactions changes the quantum tunneling probability.

We change the quantum tunneling probability in the 1-dimensional hydrodynamic simulation of the Sun to obtain the parameter range of non-standard interactions that can change the temperature of the core within allowed limits. One of the most significant systematic uncertainties that limits our analysis is the fact that the initial metallicity of the Sun has an uncertainty stemming from a disagreement between the solar metallicity values obtained from helioseismology and those inferred from the Standard Solar Model. A narrowing of the uncertainty in the solar metallicity values would lead to improved limits on the non-standard interactions.

We note that the introduced NSI will modify the Coulomb barriers not only in the key first step of the $pp$-chain. We have investigated the variability of our results by changing the Coulomb barriers, not only in the $pp$ reaction but also in the CNO bottleneck. {This was done to mimic the experimental uncertainty in that reaction cross section.} While the core temperature changed insignificantly relative to the default case, we noticed a significant alteration of the energy generation ratio from the CNO to $pp$. However, the modifications introduced by us to the $^{14}\mathrm{N} (\mathrm{p}, \gamma) ^{15}\mathrm{O}$ reaction rate are subdominant {or comparable} with respect to the SM nuclear physics experimental uncertainties. 
{
We stress that the goal of this paper is not to conduct a systematic study of the impact of experimental uncertainties in the aforementioned reactions on the solar evolution; it is rather to bring to attention that even though the fitted and extrapolated cross sections had already achieved a great level of accuracy $\mathcal{O}{(5-10\%)}$~(see reviews on the topic: Refs.~\cite{Adelberger:2010qa,Villante:2021ubp}), there is still an open door for NSI to sneak in.}
Our work thereby highlights the necessity for a better, higher sensitivity, probes of those reactions in the laboratory experiments.

There are other astrophysical systems where hydrogen fusion occurs via $pp$-chain reactions and the effects of new mediators on these systems merits further study. However, the Sun offers a wide range of observables, with a degree of statistical uncertainty that is not possible in the case of any other astrophysical system. In any case, the connection between a frontier issue in BSM particle and nuclear physics and the known properties of our nearby star is intriguing.

\acknowledgments
We would like to thank Baha Balantekin for helpful discussions. We are also grateful to Yasaman Farzan and Edoardo Vitagliano for valuable comments. AS would like to thank Irene Tamborra for the support and encouragement during the course of the project. AS and SS are supported by the Villum Foundation (Project No. 13164). GMF acknowledges NSF Grant No. PHY-1914242 at UCSD and the NSF N3AS Physics Frontier Center, NSF Grant No. PHY-2020275, and the Heising-Simons Foundation (2017-228).

\bibliographystyle{JHEP}
\bibliography{SolarNSI}
\end{document}